# Optimal Technical and Economical Operation of Microgrids through the Implementation of Sequential Quadratic Programming Algorithm

Ali Montazeri, Alireza Sedighi Anaraki, *Member IEEE*, Shahab Aref

*Abstract*-- In this paper, the optimal operation of a microgrid is investigated when the network is connected at a 24-hour interval for a specific day. For the optimal operation, some criteria, such as the simultaneous reduction in operation costs, losses, and voltage deviations, as well as the increase in reliability are taken into consideration. Given the capacity of the installed units, a part of the load of the demanded energy is always supplied by the grid. In line with the objectives set, planning for the charge and discharge of storage systems as well as the interaction among distributed power generators are investigated. The operation process is considered as an optimization problem, and the created problem is solved by the Sequential Quadratic Programming (SQP) algorithm. In addition, by implementing the Demand Response (DR) program in the optimal operation, the acquired results are compared with those obtained when this strategy is not employed. Simulation results indicate that the utilization of the Demand Response (DR) program in the optimal operation of the microgrid leads to a number of improvements. To elucidate it more, in addition to the decrease in the operation costs of the microgrid, the voltage deviation indices, reliability, and losses are improved contrary to non-DR schemes.

*Index Terms*-- Optimal operation; Microgrid, Storage, Distributed generation resources, Demand respond, Sequential Quadratic Programming (SQP).

## I. Introduction

NOWADAYS, many researchers and countries pay special attention to the utilization of renewable energies because fossil fuels are about to run out. The utilization of renewable resources in different forms leads to a decrease in losses, generation costs, and pollution, an increase in system reliability, and an improvement in voltage deviations [1]. Owing to the widespread use of distributed generation units in electric distribution systems, it is vital to take microgrids into account. The proper implementation of microgrids leads to less losses and environmental pollution in one hand, with fewer transmission lines being required to be built, on the other hand [2,3]. The main problem of renewable distributed generation units in electric distribution systems is their uncertainty, which is resolved well by energy storage units [4]. Due to the high cost of energy storage units, combining them with demand-side management will result in more reliability [5]. With the demand side response not taken into account, loads are required to be supplied fully at any moment of the day with any magnitudes, regardless of the power system efficiency. However, new methods reduce load variations to improve power system efficiency as much as possible [6]. All microgrids are required to possess a central management part to monitor loads and generation units so as to make appropriate decisions about the restrictions and goals of the power system operation [7]. Since a microgrid has a nonlinear and discrete identity, the presence of a suitable optimization algorithm is required to come up with a satisfactory solution [5]. The benefits of a central supervising unit for devising a coordinated scheme have been discussed [8]. Genetic algorithm (GA) is an optimization algorithm employed [9] to establish a smart energy management system for the operation of a power system. In order of minimizing the operation costs of microgrids, linear planning (LP) has been employed to optimize the charge and discharge of energy storage units [10]. In addition, the role of a management center aimed at optimizing operations has been investigated by considering power exchanges with an upstream power network [11]. Particle swarm and LP optimization methods could be employed to optimize the operation costs of a microgrid with energy storage units [12]. In the same vein, the energy management of a microgrid can be optimized by the Adaptive Modified Particle Swarm Optimization (AMPSO) method [13]. The Honey Bee Mating Optimization (HBMO) method can be used to optimize the energy management of a power distribution system, which contains fuel cells, wind, and solar generation units [14]. The optimal management strategy of distributed batteries in distribution systems with a high level of solar energy penetration has been discussed [15]. Demand side response can be used to reduce the losses of distribution systems [16]. Various optimizing algorithms [17], such as genetic algorithm [18, 9], particle swarm optimization [19], and an algorithm based on differential evolutions [20] are employed to enhance the operation of microgrids.

The amount of the power dedicated to each distributed generation unit and its variations in accordance with each hour of the day have direct impacts on the losses of a microgrid. Therefore, the consideration of losses in load flow calculations is vital. This paper aims to minimize the operation costs of

The authors are with Department of Electrical Engineering, Yazd University, Iran (e-mail: montazeriali@stu.yazd.ac.ir; sedighi@yazd.ac.ir; shahabaref@stu.yazd.ac.ir)

microgrids, improve their reliability, reduce losses, and modify voltage deviations. The contribution level of each distributed generation unit, the power exchange circumstance with the upstream network, the charge and discharge planning of energy storage units, as well as the losses of microgrids are considered in this paper to optimize the operation of microgrids. The objective functions are scrutinized exclusively, and their coefficients are determined by their importance level. In order to optimize the objective function, GA and Sequential Quadratic Programming (SQP) are employed, which are appropriate for the numerical solution of nonlinear optimization problems. The mathematical model of the problem is presented in section II, and the solution algorithm is explained in section III. Section IV describes the studied system, and in the end, the simulation algorithm and the results are presented in section V.

## II. MATHEMATICAL MODELING

To present a mathematical model for a microgrid connected to a power distribution network, it is essential to consider the constraints of each equipment utilized in the microgrid. The constraints must be defined so that the final optimization problem would be simple and practical. In this paper, the objective function is comprised of four main parts, with each of which having different features. However, each part is normalized and then converted into a single function by employing weighting coefficients. It is worth noting that the model presented in this paper is considered for a specific day and could be applicable to all other days of the year. Since the model presented is comprehensive, it could be utilized in other case studies.

### A. Operation Cost Function

An operation cost function consists of the cost of power generation by distributed power generation resources as well as the cost of utilizing batteries and the energy received from the upstream distribution network, which is modeled by (1), (2) [13,15] as follows:

$$F_1(P) = \sum_{t=1}^{T} \left( \sum_{i=1}^{N} B_i(P_{g_i}(t)) + P_{Grid}(t).B_{Ggrid}(t) + P_S(t).B_S \right) \quad (1)$$

$$B_i(P_{g_i}(t)) = b_i.P_{g_i}(t) + C_i \quad (2)$$

where T represents the duration considered for the investigation, N denotes the number of active distributed power generation resources in the network, $P_{gi}$ denotes the active generated power of the i[th] power source in terms of kW, and $P_{Grid}(t)$ is the active power received from the network in the t[th] hour in terms of kW. In addition, $P_s(t)$ indicates the active power of the charge and discharge of the energy storage system in the network at the t[th] hour in terms of kW. $B_{Grid}(t)$ implies the price of the energy per hour in the network in €ct. $B_S$ represents the coefficient of the cost function of using batteries in terms of €ct/kWh. $B_i(P_{gi}(t))$ is the offered price for the i[th] distributed power generation unit for generating $P_{gi}$ kilowatt of power in the t[th] hour in terms of €ct. In the end, $c_i$ and $b_i$ are the coefficients of the cost function for the i[th] unit in terms of €ct/kWh and €ct/h as illustrated in Table I.

TABLE I
The information of active distributed generation unites in microgrids [9].

| Type | $P_{min}$ (Kw) | $P_{max}$ (Kw) | $B_i$ (€ct/kwh) | $C_i$ (€ct/h) |
|------|------|------|------|------|
| PV1 | 0 | 3 | 54.84 | 0 |
| PV2 | 0 | 10 | 54.84 | 0 |
| WT | 0 | 15 | 10.63 | 0 |
| MT | 6 | 30 | 4.37 | 85.06 |
| FC | 3 | 30 | 2.86 | 255.8 |

### B. Loss function

Economic consequences and heavy costs paid for power losses have made this topic be regarded more than ever in today's electrical engineering community of the world. It is axiomatic that the amount of the power loss of the network is influenced by the amount of the distributed generation sources and batteries. One of other main objectives of this paper is to minimize the power loss according to the optimal dispatch of the distribution generation power, which is stated by (3) [21] as follows:

$$F_2(P) = \sum_{T=1}^{24} \sum_{i=1}^{N_r} (I_{i,T}^2 R_i) \quad (3)$$

where $R_i$ represents the branch resistor of B, $I^2_{i,T}$ is the current of branch B at time T, $N_r$ denotes the total number of the branches, and $F_3$ is the entire loss of the whole network in an entire 24-hour day in terms of kW.

It is evident that the value of the current of the branches is dependent on the power of DGs, batteries, and the strength of the interaction with the upstream network.

### C. Cost Function of Unsupplied Energy

One of the main purposes of the optimal operation in the microgrid considered is to improve reliability. To evaluate system reliability, there exist various indices, among which the cost of unsupplied energy is chosen, which occurs according to the error occurrence of the HV/MV transformer or MV feeders. After the occurrence of the error, the faulty equipment is separated from the network by opening some switches which are closed normally. Next, the disconnected loads are supplied by the utilization of DGs and island-made batteries in case such utilization is technically feasible. Therefore, the power distribution of DGs and the state of the charge (SOC) in batteries must be in such a way that the unsupplied energy, as one of the objective functions, would be minimized. The objective function of the unsupplied energy is stated by (4) [22, 23]:

$$F_3(P) = \sum_{t=1}^{24} \left[ \sum_{j \in A_{EV}} OC \times \lambda_j \times \left[ S_j^{OUT}(t) - S_j^{RDG}(t) - S_j^{RST}(t) \right] \times \Delta T \right] \quad (4)$$

where $F_3$ represents the function of the unsupplied energy, OC is the outage cost, $\lambda_j$ denotes the failure rate of the j[th] occurrence in each hour, and $S_j^{OUT}(t)$ denotes the power of the load disconnected by the j[th] occurrence at hour t. In addition, $S_j^{RDG}(t)$ indicates the restored power by the DG units after the j[th] occurrence in hour t, $S_j^{RST}(t)$ represents the restored power by the batteries after the j[th] occurrence in hour t, and $\Delta t$ is the time required for either repairing or replacing the equipment.

In (4), the OC value is a function of the disconnected load type and the duration of the outage. This determines the type and amount of the supplied load by means of DGs at the time of the occurrence of the error and the islanding of a part of the network [23].

*D. Voltage Deviation Function*

One of the objective functions considered is the voltage deviation function, which is modeled by (5) as [24]:

$$F_4(P) = \sum_{T=1}^{24} \sum_{i=1}^{N_r} |V_{ref} - V_{i,T}| \quad (5)$$

where $V_{ref}$ is the reference voltage being equal to 1, and $V_{i,T}$ denotes the voltage value for load points at time T. It is obvious that the voltage value at the load points is dependent on the power of DGs and batteries.

*E. System Constraints and Restrictions*

- **Power balance**

System constraints include the power balance constraint in busbars and the technical constraints related to each equipment employed in the microgrid. In (6), the power balance constraint is observed in the busbar connected to the main network. Considering this constraint, the constraint of DG generation planning and management of energy exchange with the distribution network (including electricity, charge and discharge of batteries in each hour, as well as losses in each hour) must be in such a way that the energy demanded by the load would be always supplied. This constraint is considered by (6), as follows:

$$\sum_{i=1}^{N} P_{g_i}(t) + P_{Grid}(t) = P_D(t) \pm P_S(t) + P_{Loss}(t) \quad (6)$$

where $P_D(t)$ represents the demanded power in each hour, and $P_S(t)$ is the power of the energy storage system, which could be marked differently according to the charge and discharge status of the batteries in each hour. Thus, when batteries are being charged, the demanded load must be increased. In the same vein, when batteries are being discharged, the demanded load must be decreased. By so doing, power generation sources will supply more loads at the charge time and less loads at the discharge time. $P_{Loss}(t)$ is the power loss of the microgrid in 24 hours, which is calculated using the backward-forward load distribution approach, at the presence of the batteries. It is clear that the amount of the power loss in the network is influenced by the values of DGs and batteries. That is the reason why (6) is a complex and non-linear constraint, which is solvable through the implementation of SQP algorithm.

- **Power Generation and Voltage Restriction**

The active power allocated to each source should fall within the permitted generation range of that source. In addition, the voltage magnitude of all busbars must fall within the allowed range. In other words, the required energy of the loads must be supplied normally. Such restrictions are expressed by (7) and (8) as [15]:

$$P_{g_i}^{min} \leq P_{g_i} \leq P_{g_i}^{max} \quad (7)$$

$$V_{min} \leq V_{i,t} \leq V_{max} \quad (8)$$

- **Storage system**

The stored energy in the batteries of a microgrid can supply uninterruptable loads when required. This necessitates the restriction that the amount of the stored energy in batteries must not be lower than a specific level. In this way, the state of charge (SOC) of batteries can be stated by (9) as [15]:

$$SOC(t) = SOC(t-1)(1-\delta(t)) + \eta_{charge} \max(0, P_B(t))\Delta t$$

$$+ \frac{1}{\eta_{discharge}} \min(0, P_B(t))\Delta t \quad (9)$$

where SOC (t) and SOC (t-1) represent the stored energy at times t and t-1, respectively. In addition, η indicates the efficiency of the charge and discharge of batteries, and δ is the inner discharge coefficient of the storage system, which is dependent on battery parameters, with some of which being stated in [25]. Besides, Δt is the duration of the time interval, which is considered one hour in this paper. The charge status of batteries in each hour as well as the amount of their charge and discharge in each hour are the two constraints stated in (10) and (11) as follows:

$$SOC_{min} \leq SOC(t) \leq SOC_{max} \quad (10)$$

$$|P_B(t)| \leq P_B^{max} \quad (11)$$

where values $SOC_{min}$, $SOC_{max}$, and $P_B^{max}$ are considered 16 kWh, 40 kWh, and 4 kWh, respectively [15].

- **Power system**

The restriction of the maximum exchangeable power between the microgrid and the global network could be modeled by (12) as follows:

$$P_{Grid}(t) \leq P_{Grid}^{max}(t) \quad (12)$$

$P_{Grid}^{max}$ is considered 300 kW [26].

## III. SIMULATION ALGORITHM

The optimal operation of microgrids has been modeled by (1) – (6) in section II. In other words, the optimal operation is considered an optimization problem, with the main purpose of which being to minimize different objective functions, while a number of technical and economic constraints are present. In this paper, a combination of the two algorithms of genetic algorithm and sequential quadratic programming (SQP) has been employed. Fig. 1 demonstrates the diagram of the problem-solving approach.

SQP, being one of the most proper, repetitive, and instrumental schemes, is a highly popular and powerful scheme used for numerical problem solving in non-linear optimization problems. Starting from a point, this algorithm reaches an acceptable response using the gradient of the provided data. It must be noted that SQP requires a reasonable initial point to reach a reliable response. If the optimization problem had no constraint or restriction, SQP would be reduced into a newton approach. Genetic algorithm (GA) is employed to reach the initial point of SQP algorithm. In fact, the output of GA, which is a relative optimal response, is utilized as the input of SQP algorithm to reach a more proper response. SQP utilizes the estimation of the Lagrange function in each repetition. In addition, it solves the problem by quadratic programming (QP). This quadratic program is a repetitive approach whose answer is offered in the $K^{th}$ reiteration of QP. The solving approach of SQP is presented as (13) [27]:

$$\text{minimize } 0.5 d^t H_k d + \nabla f(x_k)^t d$$
$$\text{subject to}: \nabla h_i(x_k)^t d + h_i(x_k) = 0, i = 1,...,p \quad (13)$$
$$\nabla g_i(x_k)^t d + g_i(x_k) = 0, i = 1,...,p$$

According to (14), the final $X_{K+1}$ is obtained in the $K^{th}$ repetition.

$$X_{k+1} = x_k + \alpha_k d_k \quad (14)$$

## IV. THE STUDIED SYSTEM

In this paper, the studied system is a low-voltage network with three feeders to supply domestic, industrial, and commercial loads. There are different types of DGs, including photovoltaic, fuel cells, wind turbines, micro turbines, and storage units in the simulated microgrid, with the structural information of which presented in [11]. The power factor of all DGs is assumed to be one, so no reactive power is either generated or consumed. Fig. 2 shows the studied microgrid.

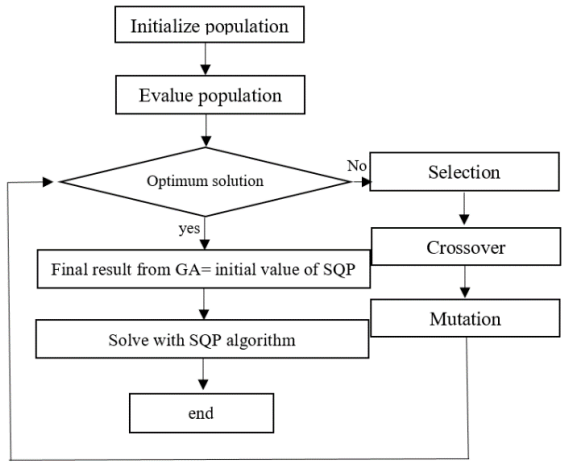

Fig. 1. Problem-solving approach

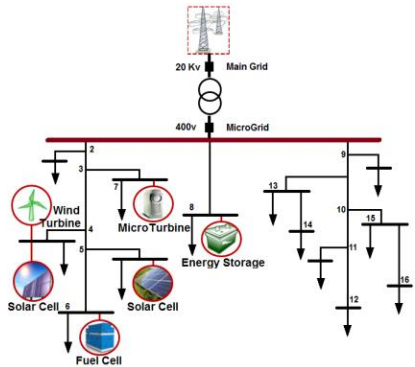

Fig. 2. The structure of studied microgrid [11].

Figs. 3 and 4 show the price of the electricity market for a 24-hour period [13] and the load variation diagram of the microgrid [11]. The generation rate of the solar and wind units are presented in [13].

## V. SIMULATION AND RESULTS

First of all, to investigate the optimal management of a microgrid, its load is supposed to be supplied by an upstream network, with all DGs and batteries being disconnected. Table II presents operation costs, losses, unsupplied energy, and voltage deviations according to the load profile of the microgrid as presented in Fig. 3. The received active power from the upstream power system is shown in Fig. 5. In addition, Fig. 6 shows the power loss of a microgrid for a 24-hour period. Next sections are related to the effectiveness of DGs, their optimum generation plan, storage systems, as well as their charge and discharge schedules, according to the optimal planning of the objective function in the operation of a microgrid. This situation is compared with the initial state in which DGs and storage systems are not considered in the network.

### A. The first scenario: the operation optimization of the microgrid to minimize operating costs

This section aims to reduce the operating costs of microgrids using the interaction among DGs, batteries, and the upstream power system. Fig. 7 shows the optimization results of this scenario. In addition, Fig. 8 shows the power exchanged among the studied network, the upstream grid, and the power loss in a 24-hour period in the initial state.

According to Fig. 8 and Table III, when scenario 1 is applied, a number of factors are decreased in comparison with the initial state, including 11.82% in costs, 22.59% in losses, 28.19% in the reliability index, and 11.04% in voltage deviations. According to Fig. 8, the optimal planning of the generation leads to a reduction in the active power exchanged with the upstream network which entails a reduction in the maximum power consumed. Based on this scenario, power losses decrease and lead to a drop in costs.

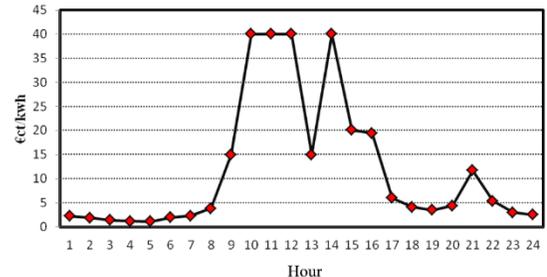

Fig 3. Proposed price of market for 24 hours [13].

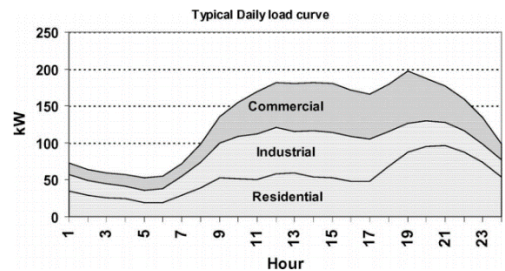

Fig. 4. Load profile of microgrid [11].

TABLE II
Results of optimum operation in initial state

| Cost (€) | 533.21 |
|---|---|
| Losses (kW) | 85.501 |
| Cost of unsupplied energy (€) | 35.05 |
| Voltage deviation | 2.2915 |

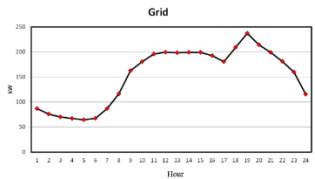

Fig. 5. Active power received from upstream network

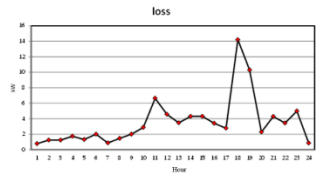

Fig. 6. Loss power of network in initial state in initial state

*B. The second scenario: the operation optimization of the microgrid to reduce power losses*

Power generation costs are excluded in this section. Hence, the contribution level of DGs, the amount of the power exchanged with the upstream network, as well as the charge and discharge plan of batteries are considered to reduce power losses in the studied system. Fig. 9 demonstrates the results of this scenario. The power exchanged between the microgrid and the upstream network as well as the 24-hour power losses of this scenario are compared with the initial state. Fig. 10 shows the results of this comparison.

According to table IV, after the implementation of the second scenario, a number of factors decreased in comparison with the initial state, including 3.6% in costs, 35.99% in losses, and 30.47% in the costs of unsupplied energy, but the voltage deviation index increased about 4.49%. According to Fig. 10, the losses and the power exchanged with the upstream network decreased significantly, which resulted in a reduction in the maximum load in the power system.

*C. The third scenario: the operation optimization of the microgrid to reduce the cost of the unsupplied energy*

This scenario is based on the fact that the required loads of the network are supplied so that the costs of unsupplied energy are minimized. Hence, the contribution level of DGs in electricity generation, power exchange with the upstream network, as well as the charge and discharge plan of batteries are considered to minimize the cost of unsupplied energy. Fig. 11 shows the results of this scenario.

According to Fig. 12 and Table V, the implementation of the third scenario decreased some factors, including 1.36% in costs, 23.9% in losses, and 34.4% in the reliability index, but voltage deviations increased by about 0.2%. The optimal generation planning reduced the power exchanged with the upstream network and resulted in a decrease in the consumed load peak

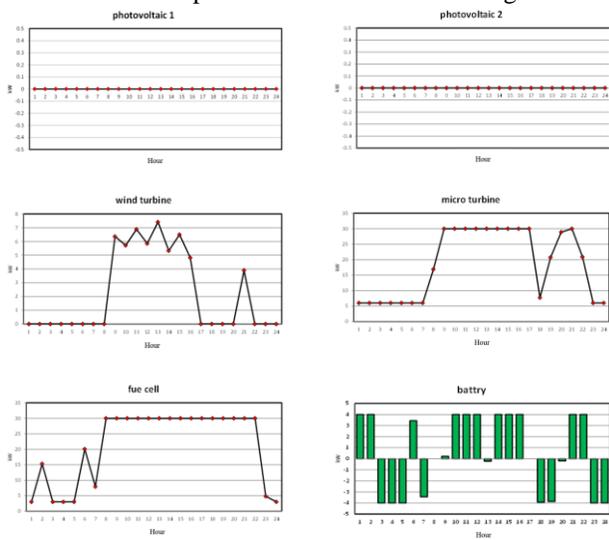

Fig. 7. Dedicated active power to DGs and charge and discharge of batteries

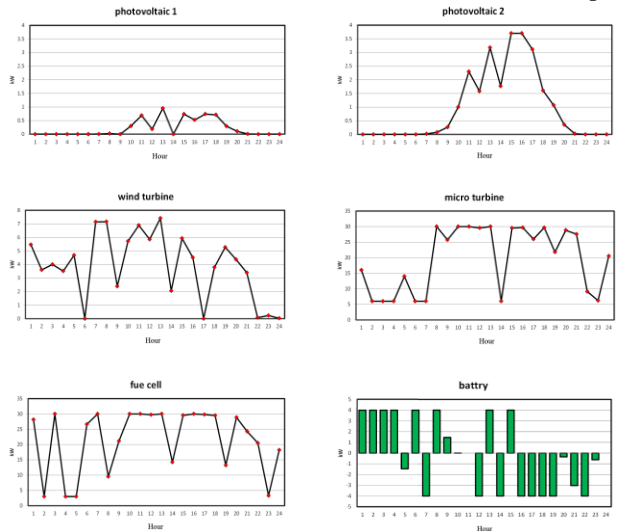

Fig. 9. Devoted power to generation units, and charge and discharge plan of batteries

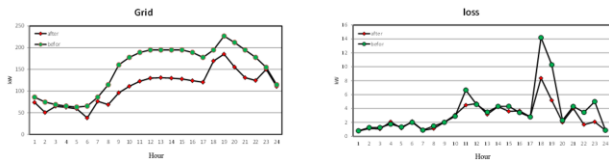

Fig. 8. Comparison between first scenario and initial state in terms of the active power exchanged between microgrid and upstream network and hourly loss power

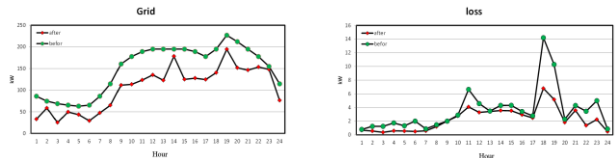

Fig. 10. Comparison between second scenario and initial state in terms of the active power exchanged between microgrid and upstream network and hourly loss power

TABLE III
Results of optimum operation due to first scenario

|  | Cost (€) | Losses (kW) | Cost of unsupplied energy (€) | Voltage deviation index |
|---|---|---|---|---|
| Initial state | 533.21 | 85.501 | 36.04 | 2.2915 |
| First scenario | 470.18 | 66.187 | 25.88 | 2.0386 |

TABLE IV
Results of optimum operation due to second scenario

|  | Cost (€) | Losses (kW) | Cost of unsupplied energy (€) | Voltage deviation index |
|---|---|---|---|---|
| Initial state | 533.21 | 85.501 | 36.04 | 2.2915 |
| Second scenario | 513.93 | 54.73 | 25.06 | 2.3944 |

of the network. The costs are reduced in this scenario due to the loss reduction.

### D. The forth scenario: the operation optimization of the microgrid to reduce voltage deviations

The main purpose of this scenario is to reduce voltage deviations. Hence, the contribution level of DGs in electricity generation, power exchanged with the upstream network, as well as the charge and discharge plan of batteries are considered to reduce voltage deviations. Fig. 13 shows the results of this scenario.

According to Fig. 14 and Table VI, the forth scenario reduced some factors in comparison with the initial state, including 0.42% in costs, 2.96% in losses, 26.8% in the reliability index, as well as 31.92% in voltage deviations. Just like previous generation planning, this scheme reduces power exchanged with the upstream network and decreases the consumed load peak of the network as well.

### E. The fifth scenario: the optimal operation of the microgrid

The purpose of the optimal operation in this section is to reduce operation costs, power losses, voltage deviations, and the cost of unsupplied energy simultaneously. In previous scenarios, each of these parameters was optimized individually. Four objective functions are combined to create the primary objective function. To this end, objective functions are normalized, and then using suitable weighting coefficients, one function is obtained from all other functions, as Eq. 15 shows:

$$F = W_1 \left( \frac{F_1 - F_{1,\min}}{F_{1,\max} - F_{1,\min}} \right) + W_2 \left( \frac{F_2 - F_{2,\min}}{F_{2,\max} - F_{2,\min}} \right) \\ + W_3 \left( \frac{F_3 - F_{3,\min}}{F_{3,\max} - F_{3,\min}} \right) + W_4 \left( \frac{F_4 - F_{4,\min}}{F_{4,\max} - F_{4,\min}} \right) \quad (15)$$

where the minimum value of each function is its singular optimal value, and the maximum value of each function is the value of the function before executing the program. It must be noted that $W_i$ denotes the importance of each function. This value is determined using an AHP process.

## VI. THE ANALYTICAL HIERARCHY PROCESS

As already noted, for the optimal operation of the micro-grid, the weighted sum of the single objective functions was used to turn the multi-objective function into a single one. To determine the weighting coefficients, different methods such as goal attainment and the fuzzy approach were used. One of the most efficient techniques for weighting multi-objective functions is the Analytical Hierarchy Process (AHP). This technique was introduced by Thomas L. Saati (1980), and it has been the focus of attention in scientific circles for the past

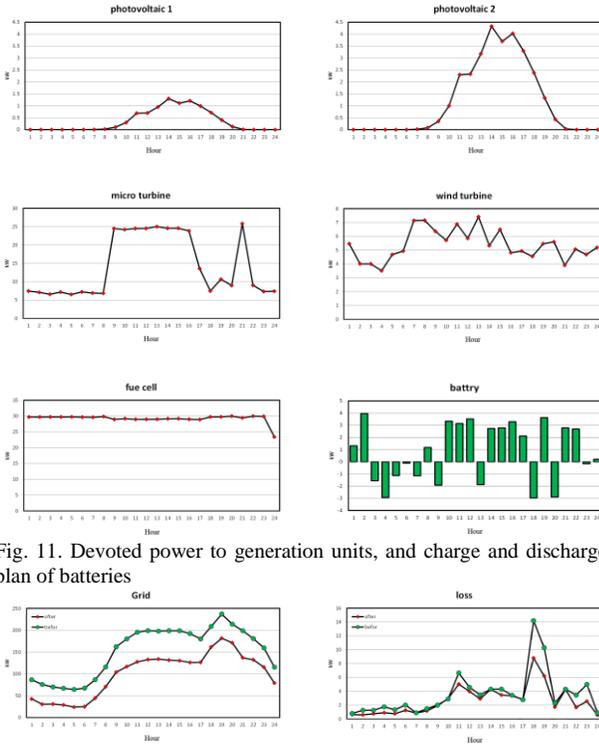

Fig. 11. Devoted power to generation units, and charge and discharge plan of batteries

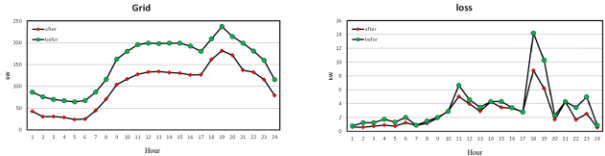

Fig. 12. Comparison between third scenario and initial state in terms of the active power exchanged between microgrid and upstream network and hourly loss power

TABLE V
Results of optimum operation due to third scenario

|  | Cost (€) | Losses (kW) | Cost of unsupplied energy (€) | Voltage deviation index |
|---|---|---|---|---|
| Initial state | 533.21 | 85.501 | 36.04 | 2.2915 |
| Third scenario | 525.95 | 65.063 | 23.64 | 2.2962 |

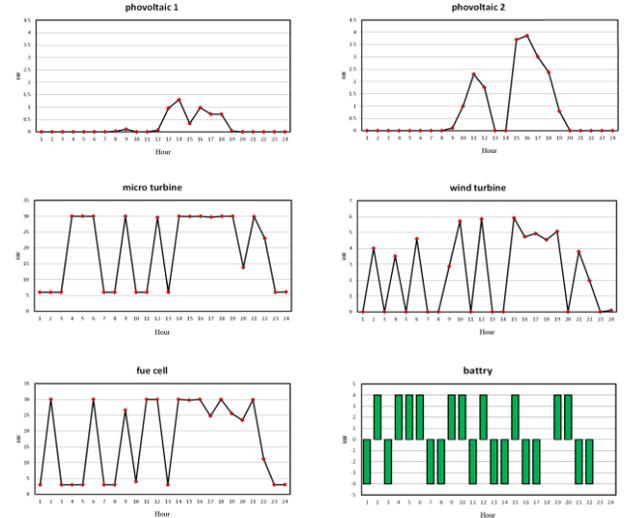

Fig. 13. Devoted power to generation units, and charge and discharge plan of batteries

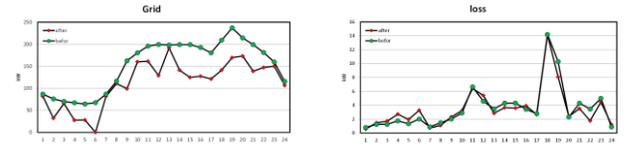

Fig. 14. Comparison between forth scenario and initial state in terms of the active power exchanged between microgrid and upstream network and hourly loss power

TABLE VI
Results of optimum operation due to forth scenario

|  | Cost (€) | Losses (kW) | Cost of unsupplied energy (€) | Voltage deviation index |
|---|---|---|---|---|
| Initial state | 533.21 | 85.501 | 36.04 | 2.2915 |
| Forth scenario | 530.95 | 82.97 | 26.38 | 1.5601 |

20 years. In this paper, the Analytical Hierarchy Process has been used to weight multi-objective functions.

The weighted coefficients of the multi-objective function were determined in accordance with the descriptions offered in [28], by the Expert Choice software.

The objective function weighting coefficients obtained by the Analytical Hierarchy Process using the Expert Choice software are presented in Eq. 16.

$$F = 0.157 \times \left(\frac{F_1 - F_{1,\min}}{F_{1,\max} - F_{1,\min}}\right) + 0.483 \times \left(\frac{F_2 - F_{2,\min}}{F_{2,\max} - F_{2,\min}}\right)$$
$$+ 0.272 \times \left(\frac{F_3 - F_{3,\min}}{F_{3,\max} - F_{3,\min}}\right) + 0.088 \times \left(\frac{F_4 - F_{4,\min}}{F_{4,\max} - F_{4,\min}}\right) \quad (16)$$

The results of this scenario are presented in Fig. 15.

According to Fig. 16 and the results of comparing scenario 5 and the base mode in Table VII, costs, losses, the reliability index, and voltage deviations are decreased by 11.21%, 27.68%, 32.05%, and 11.89%, respectively, by applying the results of scenario 5. In addition, according to Fig. 16, optimal production scheduling reduces the power exchanged with the upstream network, thereby resulting in a reduction in the total network consumption peak.

Load management is a method employed by system operators, in which they encourage consumers to reduce their load at peak or expensive times and shift consumption to off-peak hours. In this method, apart from reducing the costs for consumers, the benefits of operating a micro-grid are increased. By applying the load demand response strategy (DR), which is a method of load management, the obtained results are compared with those of the case in which this strategy is not applied to optimize the operation of the micro-grid. Fig. 17 shows the results obtained from applying consumed load management to the optimal operation.

As Table VIII shows, by the simultaneous applying of load management to the optimal operation, it is observed that all parameters studied in this paper are reduced significantly.

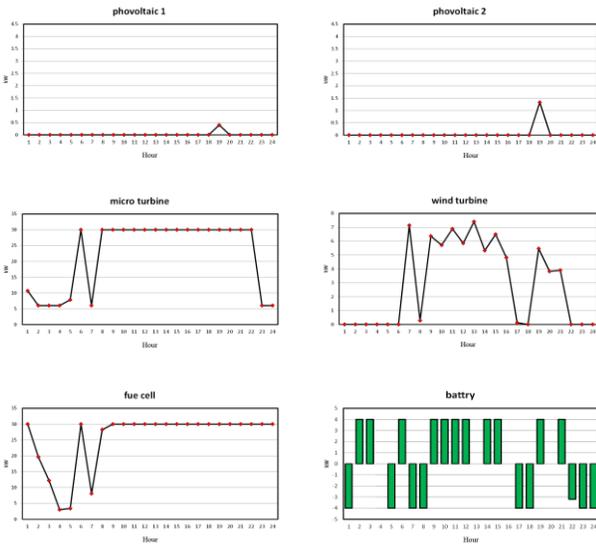

Fig. 15. Devoted power to generation units, and charge and discharge plan of batteries

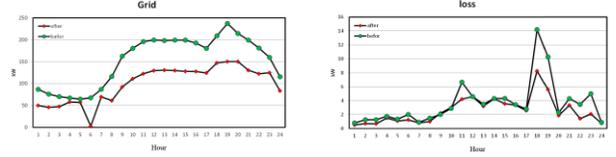

Fig. 16. Comparison between first scenario and initial state in terms of the active power exchanged between microgrid and upstream network and hourly loss power

TABLE VII
Results of optimum operation due to fifth scenario

|  | Cost (€) | Losses (kW) | Cost of unsupplied energy (€) | Voltage deviation index |
|---|---|---|---|---|
| Initial state | 533.21 | 85.501 | 36.04 | 2.2915 |
| Fifth scenario | 473.46 | 61.588 | 24.49 | 2.0191 |

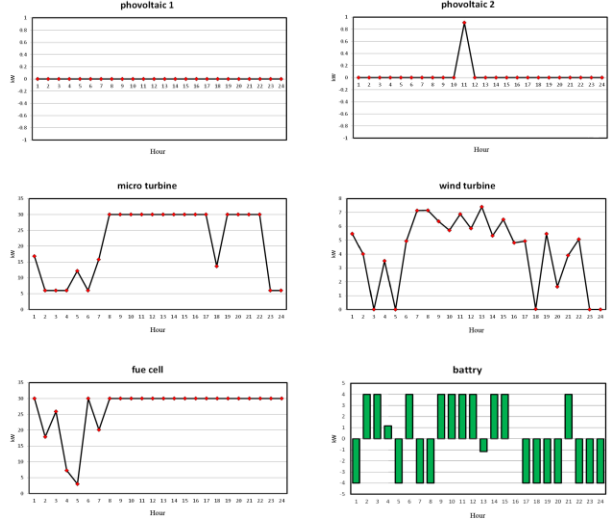

Fig. 17. Devoted power to generation units, and charge and discharge plan of batteries

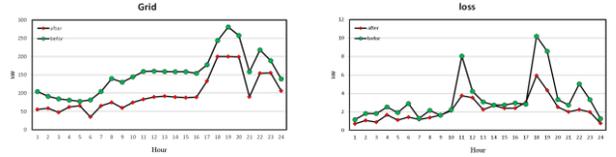

Fig. 18. Comparison between active power exchanged between microgrid and upstream network and hourly loss power

TABLE VIII
Comparison between optimum operation results

|  | Cost (€) | Losses (kW) | Cost of unsupplied energy (€) | Voltage deviation index |
|---|---|---|---|---|
| Initial state | 533.21 | 85.501 | 36.04 | 2.2915 |
| Fifth scenario without DR | 473.46 | 61.588 | 24.49 | 2.0191 |
| Fifth scenario with DR | 393.95 | 52.67 | 24.17 | 1.9673 |

Figs. 19 illustrate the power exchanged with the grid and the consumed load before and after DR implementation.

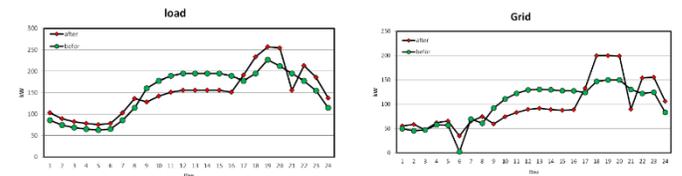

Fig. 19. Power exchanged with the grid before and after DR and The consumed load of grid before and after DR

## VII. CONCLUSION

In this paper, after modeling objective functions and introducing the studied system, the optimal operation of microgrids was examined at the presence of distributed generation sources as well as the storage system in various scenarios. In this study, both the amount of cooperation and the way batteries are charged and discharged are analyzed, with the operation duration of the microgrid considered to be 24 hours. SQP was the utilized algorithm, being a fast algorithm in reaching the final solution. An initial supposed value was also determined to solve the problem by Genetic algorithm in a way that the convergence of the solution would occur more quickly. In addition, the results obtained from the optimization process, including the way of supplying the load demand in the distributed generation source, the scheduling of the charge and discharge of batteries, and the received power from the distribution network were presented. The results were compared with conditions before utilizing optimal programming. In addition, the impact of using DR was investigated. According to the results obtained, the simultaneous implementation of optimal programming and the demand response (DR) culminated in interesting results; in other words, during the low loading time intervals, when the expense of purchasing electricity from the upstream network is low, the required power of the microgrid is mostly acquired from the upstream network. In addition, during the peak load period when the price of purchasing electricity from the upstream network is high, distributed generation sources play the major role in supplying the required load of the network. Batteries are charged and discharged in a way that when acting addition, in the period of the peak load when electricity is expensive, the power of batteries is injected into the network to decrease the operating price in the whole microgrid.